\documentclass[aps,prl,twocolumn]{revtex4-1}
\usepackage{mathrsfs}
\usepackage{physics}
\usepackage{hyperref}
\usepackage{bm}
\usepackage{url} 
\usepackage{graphicx}
\usepackage{subfigure}
\usepackage{xcolor}
\usepackage{bm}
\usepackage{bbm}
\usepackage{times}
\usepackage{stix}
\usepackage{soul}
\newcommand{\ba}{\begin{equation}\begin{aligned}}
\newcommand{\ea}{\end{aligned}\end{equation}}

\newcommand{\mm}[1]{\bm{\mathit{#1}}} 
\newcommand{\W}{\mathcal{W}}

\begin{document}
\title{Dynamical Phase Transitions in a 2D Classical Nonequilibrium Model via 2D Tensor Networks}
\author{Phillip Helms}
\email{phelms@caltech.edu}
\author{Garnet Kin-Lic Chan}
\email{garnetc@caltech.edu}
\affiliation{Division of Chemistry and Chemical Engineering, California Institute of Technology, Pasadena,
CA 91125}
\date{\today}

\begin{abstract}
  We demonstrate the power of 2D tensor networks for obtaining large deviation functions
  of dynamical observables in a classical nonequilibrium setting. 
Using these methods, we analyze the previously unstudied dynamical phase behavior of the 
fully 2D asymmetric simple exclusion process with biases in both the $x$ and $y$ directions.
We identify a dynamical phase transition, from a jammed to a flowing phase, 
and characterize the phases and the transition, with an estimate of the critical point and exponents.
\end{abstract}

\maketitle


\noindent
{\textbf{\emph{Introduction --}}}
Large deviation theory (LDT) has emerged as a powerful framework for studying
the fluctuations of macroscopic dynamical observables in classical nonequilibrium systems
~\cite{derrida2007non,garrahan2009first,prados2011large,ray2018importance,touchette2009large}. 
Reminiscent of equilibrium statistical mechanics, where ensembles of configurations are organized by their 
macroscopic properties, such as temperature or energy, LDT prescribes the grouping of 
trajectories into ensembles based on their dynamical or static macroscopic properties, such as current or density. 
This approach allows for the definition of dynamical partition functions, derivatives of which are
the mathematical analogs to entropy and free energy, named large deviation functions (LDFs),
which encode the statistics of dynamical observable fluctuations. 
As in equilibrium systems, these are critical for identifying and characterizing phase transitions, 
particularly those which occur in the space of trajectories, called dynamical phase transitions (DPTs)~\cite{touchette2009large}.

The success of LDT has been accompanied by the development of
numerical methods for computing LDFs, with significant emphasis and progress 
centered in sophisticated sampling techniques
~\cite{ray2018importance,ray2018exact,nemoto2016population,klymko2018rare,nemoto2017finite,ray2019constructing,margazoglou2019hybrid,das2019variational}.
Alternatively, the matrix product ansatz, a powerful analytical representation of nonequilibrium steady states
~\cite{blythe2007nonequilibrium,prolhac2009matrix,derrida1993exact},
foreshadowed the recent success of numerical tensor network (TN) algorithms. 
In particular, calculations using matrix product states (MPS), the 1D TN that underpins
the density matrix renormalization group (DMRG) algorithm~\cite{schollwock2011density}, provide a noiseless 
alternative to sampling methods. 
As demonstrated in the recent applications to DPTs in kinetically constrained and driven diffusive models
~\cite{gorissen2012exact,gorissen2011finite,gorissen2009density,banuls2019using,helms2019dynamical}, the MPS provides a
remarkably compact representation of nonequilibrium steady states.

While the TN approach is promising, the use of the MPS, which only efficiently encodes correlations in one dimension, 
limits the study of higher dimensional problems~\cite{stoudenmire2012studying}, 
Consequently, LDF computations beyond one dimension have relied on Monte Carlo methods
~\cite{tizon2017order,garrahan2007dynamical,hedges2009dynamic,chandler2010dynamics}.
In this letter, we demonstrate how an inherently 2D TN, the projected entangled pair state (PEPS)
~\cite{verstraete2006criticality,lubasch2014unifying,orus2014practical,phien2015infinite},
serves as an efficient ansatz to determine LDFs in 2D nonequilibrium lattice problems.

We use this approach to obtain new insights into the fully 2D
asymmetric simple exclusion process (ASEP). 
In 1D, the ASEP has become a paradigmatic model of nonequilibrium behavior
frequently employed to understand important physical systems and phenomena including
surface growth~\cite{krug1997origins,odor2009mapping}, 
molecular motors~\cite{klumpp2003traffic,chou2004clustered,lipowsky2006molecular}, 
and traffic flow~\cite{schadschneider2000statistical}.
The 2D ASEP is of similarly wide interest, but it has remained poorly characterized
~\cite{alexander1992shock,ding2018analytical,singh2009transverse,yau2004law,tamm2010statistics,schmittmann1992onset},
especially with regards to its dynamical phase behavior, which is unknown except in the periodic, weakly asymmetric limit
~\cite{tizon2017order}. We show that 2D TN now allow us to shed light on the general 2D ASEP,
by computing detailed observables along a line in the dynamical phase diagram. In so doing, we find and characterize a hitherto unobserved
DPT between jammed and flowing phases. 

\smallskip 
\noindent
{\textbf{\emph{Large Deviation Theory and Projected Entangled Pair States --}}}
We begin with a short overview of relevant theory and methods associated with 
LDT, TNs, and PEPS. More comprehensive treatments of all three topics are provided in 
recent reviews and methodological papers~\cite{touchette2009large,lubasch2014unifying,phien2015infinite}.

\begin{figure}
\centering
\includegraphics[scale=0.32]{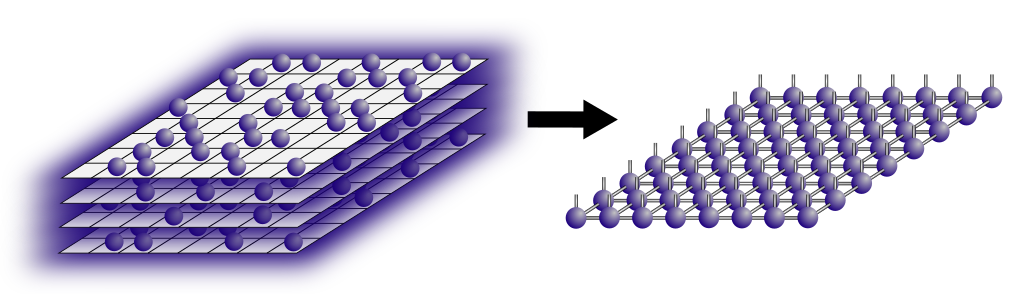}
\caption[ASEP and PEPS]{
A stack of possible configurations of the 2D ASEP (left), 
representing all possible configuration probabilities, 
is stored as a 2D PEPS, whose TN diagram is shown on the right. 
Contracting all auxiliary bonds gives the probability of all possible lattice configurations.
}
\label{fig:peps}
\end{figure}

\begin{figure*}[ht!]
\centering
\includegraphics{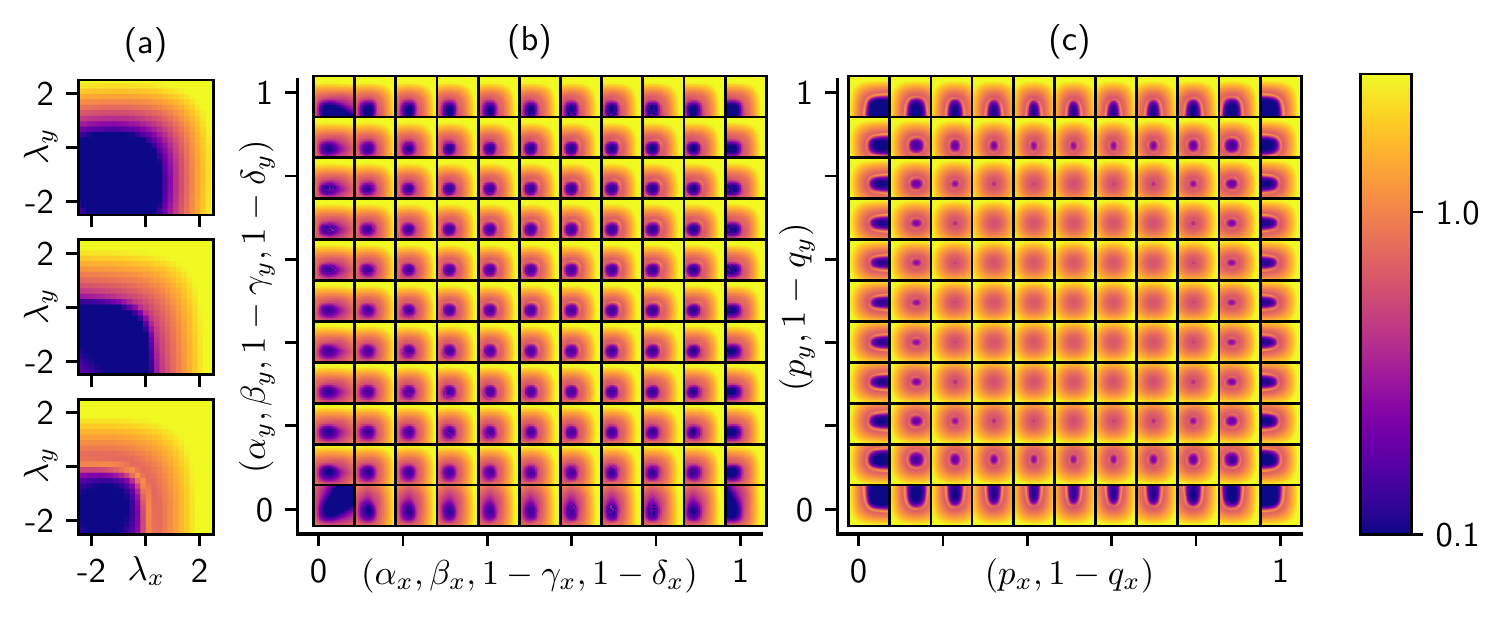}
\caption[Mean Field Dynamical Phase Diagram]{
A mapping of the mean field dynamical phase diagram of the 2D ASEP with 
(a) showing the SCGF (top), current (middle), and current susceptibility (bottom) 
as a function of bias at one point in the physical phase space,
while (b) and (c) respectively show plots of the current susceptibility as a function of bias for a bulk biased
and a boundary biased 2D ASEP. 
For (a), $p_{x,y}=1-q_{x,y}=1$ with boundary terms at $1/2$ and current biases,
$\lambda_x,\lambda_y\in[-2.5,2.5]$; we can see the transition between the jammed (dark) and flowing (bright) phases.
In (b), bulk rates are fixed at $p_{x,y}=1-q_{x,y}=0.9$ 
while sweeping over a subset of boundary rates 
($\alpha_{x,y}=\beta_{x,y}=1-\gamma_{x,y}=1-\delta_{x,y}$).
In (c), all boundary terms are set to $1/2$
and we sweep over bulk hopping rates ($p_{x,y}$,$q_{x,y}$).
Each subplot in (b) and (c) sweeps over current biases
$\lambda_x,\lambda_y\in[-2.5,2.5]$.}
\label{fig:mfdpd}
\end{figure*}

A Markovian nonequilibrium system's time evolution is governed by a master equation,
$\partial_t|P_t\rangle=\mm{\W}|P_t\rangle$,
where vector $|P_t\rangle$ represents the configurational probabilities at time $t$ 
and the generator, $\mm{\W}$, dictates the transition rates between configurations. 
At steady-state, the time-averaged current vector, $\bar{\mm{J}}=\mm{J}/t$
obeys a large deviation principle, $P(\bar{\mm{J}})\approx e^{-t \phi(\bar{\mm{J}})}$, as does 
its moment generating function, $Z(\mm{\lambda}) = \langle e^{-\mm{\lambda} \bar{\mm{J}}}\rangle\approx e^{-t\psi(\mm{\lambda})}$,
indicating that the probability of observing all but the most likely current decays exponentially with averaging time.
The rate function (RF), $\phi(\bar{\mm{J}})$, defines the probability of a given current, and $\psi(\mm{\lambda})$ is the
scaled cumulant generating function (SCGF), whose derivatives at $\mm{\lambda}=0$ give the cumulants of the current.

Performing a tilting of the generator, $\mm{\W}\to \mm{\W}^{(\mm{\lambda})}$, effectively weights trajectories according to their currents,
by scaling all forward (backward) hopping terms by $e^{-\mm{\lambda}}$ ($e^{\mm{\lambda}}$), 
making $\mm{\W}^{(\mm{\lambda})}$ non-Markovian and non-Hermitian.
A central finding in LDT dictates that the largest eigenvalue of the tilted generator is the SCGF, 
i.e. $\mm{\W}^{(\mm{\lambda})}|P^{(\mm{\lambda})}\rangle = \psi(\mm{\lambda})|P^{(\mm{\lambda})}\rangle$.
Furthermore, the corresponding left and right eigenvectors detail trajectory characteristics associated with particular fluctuations.
For example, the time averaged local density associated with a fluctuation is 
$\rho_i = \langle P^{(\mm{\lambda})}|n_i|P^{(\mm{\lambda})}\rangle/\langle P^{(\mm{\lambda})}|P^{(\mm{\lambda})}\rangle$, 
where $n_i$ is the particle number operator acting on site $i$ and $\langle P^{(\mm{\lambda})}|$ and $|P^{(\mm{\lambda})}\rangle$ are the left and right eigenvectors.

The PEPS TN ansatz is a intuitive representation of the approximate eigenstates of the tilted generator and
a diagrammatic representation of this ansatz is shown on the right side of Figure \ref{fig:peps},
where a tensor is allocated for each lattice site.
Diagrammatically, each tensor is represented as a ball with tensor indices corresponding to lines connected to the ball.
The vertical indices, called the physical bonds, correspond to the local state space of the system 
and are of size $d$, which is the local state dimension
(for hard core particles $d=2$, corresponding to an empty or occupied site).
Additionally, nearest neighbor tensors are connected by indices, called auxiliary bonds, of size $D$, 
enabling information transfer between sites.
This results in a lattice of rank five bulk tensors $\mathcal{T}^{[x,y]}_{ijklm}$ of size $(d,D,D,D,D)$.
The size of the auxiliary bonds, called the bond dimension, controls the accuracy of the ansatz 
by truncating the considered Hilbert space and for sufficiently large $D$ the ansatz is exact.
While $D$ must grow exponentially with the size of the lattice to accurately represent arbitrary states,
in practice, many states are accurately captured by a PEPS with finite $D$ even as the lattice grows.  
By contracting over all auxiliary bonds, the eigenstate of the tilted generator is recovered,
thus the mapping in Figure \ref{fig:peps} roughly illustrates how the set of all configurational probabilities
are stored as a PEPS.

The development of appropriate PEPS optimization methods for 
quantum many body problems is an active area of research
~\cite{corboz2016variational,vanderstraeten2019simulating,o2019simplified,haghshenas2019conversion}.
For this work, we simply adapt many of the most successful standard techniques to the non-equilibrium master equation setting.
Using the time-evolving block decimation approach~\cite{lubasch2014algorithms,lubasch2014unifying}, 
we integrate the tilted master equation forwards in time, giving 
$|P^{(\mm{\lambda})}_t\rangle=e^{t\mm{\W}^{(\mm{\lambda})}}|P^{(\mm{\lambda})}_0\rangle$.
We apply the time evolution operator to the initial PEPS via its
 Suzuki-Trotter decomposition into local gates, 
 $e^{t\mm{\W}^{(\mm{\lambda})}}\approx \left(e^{\delta t\mm{\W}^{(\mm{\lambda})}_{i,i+1}}\right)^{t/\delta t}$,
 and iterate this application until convergence to the steady-state.
The bond dimension between two sites grows after the application of the gate, 
thus an alternating least squares approach is used to compress the tensors
back to dimension $D$~\cite{phien2015infinite}. 
The alternating least squares algorithm uses information from all the other tensors
which are contracted into an approximate environment
using the single-layer boundary method~\cite{PhysRevB.96.045128}
and tensor reduction~\cite{PhysRevB.81.165104,PhysRevA.81.050303}.
The accuracy of the environment is then determined by an additional parameter, $\chi$,
which corresponds to the bond dimension of a boundary MPS. Like $D$, $\chi$ must
also be increased to converge to the exact stationary state.
In practice, because the environment computation is expensive, we can first
determine an approximate stationary state via the ``simple update'' algorithm where
no environment is used~\cite{jiang2008accurate}; 
then $D$ and $\chi$ are increased in subsequent time evolution steps using the full environment information (``full update'' algorithm~\cite{lubasch2014algorithms})
while $\delta t$ is also decreased to reduce the Suzuki-Trotter error.

\begin{figure*}[t]
\centering
\includegraphics{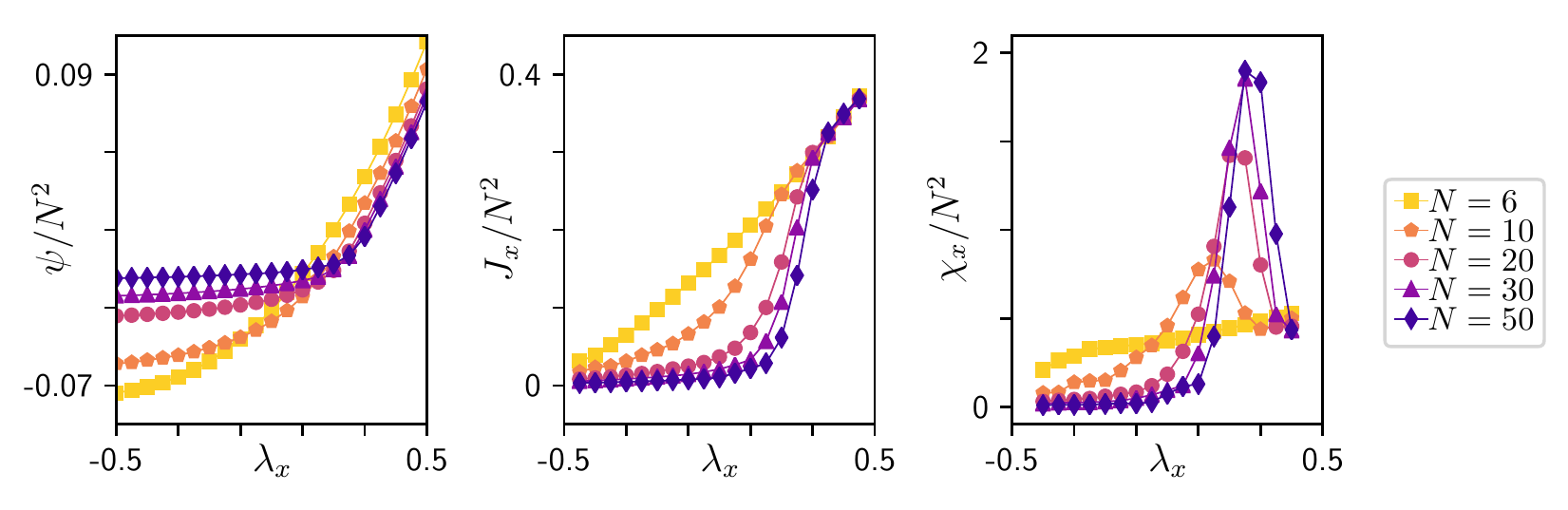}
\caption[Dynamical Phase Transition]{
PEPS calculation results analyzing the phase transition along a line in the dynamical phase space of the 2D ASEP. 
From left to right, we show the per site SCGF $\psi(\lambda_x,\lambda_y)/N^2$, horizontal current $J_x/N^2$, 
and horizontal current susceptibility $\chi_x/N^2$
at $\lambda_y=-1/2$ with $\lambda_x\in[-1/2,1]$. 
Each line corresponds to a system size $N\in[6,10,20,30,50]$.
}
\label{fig:dpt}
\end{figure*}

\smallskip 
\noindent
{\textbf{\emph{Model: 2D ASEP --}}}
The 2D ASEP, Figure \ref{fig:peps} (left), takes place on a square $N\times N$ lattice, 
where each site may be occupied by a particle or empty.
Particles stochastically hop into vacant nearest-neighbor lattice sites in the right (up) and left (down) 
directions at rates $p_x$ ($p_y$) and $q_x$ ($q_y$) respectively. 
At the $\{\text{left},\text{bottom},\text{right},\text{top}\}$ boundaries, particles are inserted at rates 
$\{\alpha_x,\alpha_y,\delta_x,\delta_y\}$, and removed at rates $\{\gamma_x,\gamma_y,\beta_x,\beta_y\}$. 
Additionally, as detailed in the previous section, we utilize a current bias in both directions,
$\mm{\lambda}=(\lambda_x,\lambda_y)$,
to probe the trajectory phase space.
The tilted generator is built from hopping operators 
$\mm{o}^{\text{hop}}_{i,j}=r_{i,j}(e^{\lambda_{i,j}}\mm{a}_i\mm{a}_{j}^\dagger - \mm{n}_i \mm{v}_{j})$ 
and similarly defined insertion and removal operators,
where $r_{i,j}$ is the hopping rate from site $i$ to $j$ and 
$\mm{a}_i$, $\mm{a}_{i}^\dagger$, $\mm{n}_i$, and $\mm{v}_{i}$ are respectively
annihilation, creation, particle number and vacancy operators. 
Because hopping occurs only between nearest neighbor sites, 
the full tilted generator, $\mm{\W}^{(\mm{\lambda})}$, then decomposes naturally into nearest neighbor gates.
At $\lambda_{i,j}=0$, $\forall (i,j)$, the system undergoes its typical dynamics, 
otherwise the biasing allows for probing of rare trajectories.


\smallskip 
\noindent
{\textbf{\emph{Results --}}}
We first probed for the existence of a DPT in the 2D ASEP by performing mean field (MF) 
computations of the SCGF on an $8\times8$ lattice in two subsets of the phase space, 
with results shown in Figure \ref{fig:mfdpd}.
In Figure \ref{fig:mfdpd}(a) we show, from top to bottom, the per site SCGF, total current, and current susceptibility 
at $p_{x,y}=1-q_{x,y}=1$ with $\alpha_{x,y}=\beta_{x,y}=\gamma_{x,y}=\delta_{x,y}=1/2$.
and current biases sweeping over $\lambda_x,\lambda_y\in[-2.5,2.5]$.
In the bottom left of these plots, we see a low-current regime materialize, where the SCGF and current flattens,
bounded by a small peak in the susceptibility 
(the thin bright line between the purple and orange regions). 

To further explore where this low-current phase materializes, Figure \ref{fig:mfdpd} (b) and (c) 
contain subplots at various points in the rate parameter space, 
each showing the per site current susceptibility as a function of $\lambda_{x,y}\in[-2.5,2.5]$.
(b) explores boundary effects, sweeping boundary terms with 
$\alpha_{x,y}=\beta_{x,y}=1-\gamma_{x,y}=1-\delta_{x,y}$ 
and maintaining asymmetric interior rates $p_{x,y}=1-q_{x,y}=0.9$
while (c) probes the effect of bulk hopping rates, 
sweeping interior hopping rates while holding boundary terms at 
$\alpha_{x,y}=\beta_{x,y}=\gamma_{x,y}=\delta_{x,y}=1/2$.

Phase transitions can be marked by a peak in the current susceptibility, as seen in Figure \ref{fig:mfdpd} (a). 
In Figure \ref{fig:mfdpd} (c) this becomes visible at sufficiently high biases ($\approx p_x>0.8$),
again accompanied by a region of distinctly low current.
This aligns with the known behavior of the 1D ASEP, where a DPT is observed except when $p_x=q_x=1/2$, 
which corresponds to the Symmetric Simple Exclusion Process (SSEP).
Furthermore, intuition from the 1D ASEP would further predict a DPT to appear for low biases in the thermodynamic limit. 
For the boundary biased results, Figure \ref{fig:mfdpd} (b), we observe the boundary rates to have little effect, 
except at extreme values, where the location of the DPT becomes distorted 
due to no insertion or removal at a boundary.


Selecting a line within the phase space covered in Figure \ref{fig:mfdpd} (c) at $p_{x,y}=1-q_{x,y}=0.9$
$\lambda_y=-1/2$ with $\lambda_x\in[-1/2,1/2]$, we carried out PEPS calculations on 
$N\times N$ lattices with $N\in\{6,10,20,30,50\}$ to probe the DPT's finite size behavior.
Here, we used $D\in[2,8]$ and $\chi=80$ while systematically reducing $\delta t\in[10^{-1},10^{-4}]$. 
Figure \ref{fig:dpt} displays key results from these calculations in support of the existence of a DPT.

There, the left plot shows the SCGF for the $\lambda_x$ sweep, 
with the flattening of the curve for large systems on the left side of the plot indicating
a low-current region.
The horizontal current $J_x$ and current susceptibility $\chi_x$, shown in the center and right plots, 
are computed via central difference numerical differentiation with respect to $\lambda_x$;
while they can also be computed via contractions with the left and right PEPS eigenstates of $\mm{\W}^{(\mm{\lambda})}$,
for the largest systems this can be numerically challenging and requires well-converged left and right states.

In all plots, we see two distinct regions, indicative of a DPT. 
Moving from right to left, we see the emergence of a low-current phase at $\approx\lambda_x=1/4$, 
where both $J_x$ and $J_y$ (not shown) are small.
The transition becomes sharper as the size of the lattice increases,
as seen by the increasingly large peaks in current susceptibility,
substantiating the existence of a second-order DPT between the jammed and flowing phases.
Furthermore, the most likely configurations in the 
flowing phase are those where particles are evenly distributed throughout the lattice, 
while in the low-current phase, those most likely are entirely filled, jamming flow in the bulk.


\begin{figure}
\centering
\includegraphics{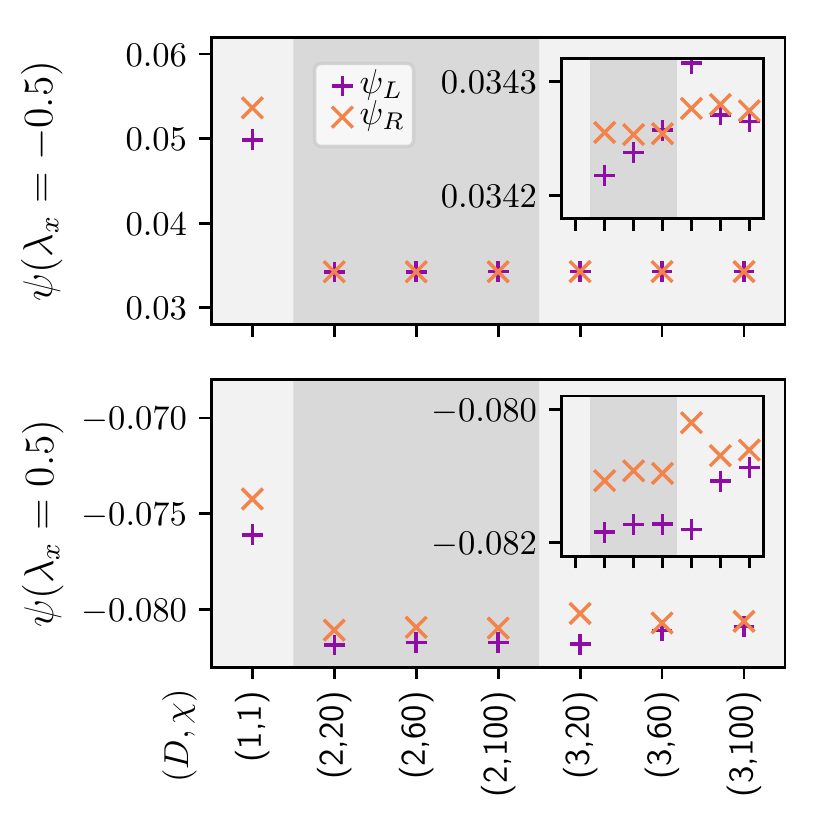}
\caption[PEPS Convergence]{
The convergence of PEPS calculations, showing the SCGF,
computed as the left and right eigenvalue of the tilted generator, $\psi_L$ and $\psi_R$,
for a $20\time20$ lattice as a function of the bond dimension $D$ (shaded) and the
boundary bond dimension $\chi$ (labeled as $(D,\chi)$. 
The top (bottom) plot corresponds to results 
in the jammed (flowing) phase at $\lambda=-0.5$ ($\lambda=0.5$).
The insets provide magnified results to illustrate the extent of convergence. 
}
\label{fig:conv}
\end{figure}

To gauge the accuracy of these results, Figure \ref{fig:conv} displays the convergence of the SCGF for 
calculations with $N=20$.
Here, the SCGF is computed from the right and left eigenstates, $\psi_R$ and $\psi_L$, 
in the jammed (top) and flowing (bottom) phases with $\lambda=-0.5$ and $\lambda=0.5$ respectively. 
Shaded regions correspond to $D$, starting with mean field results on the left and increasing to the right, 
where within each shaded region, the accuracy is improved by increasing $\chi$. 
Each computation was performed independently, doing the ''full update'' procedure from a
random initial state, decreasing the time step sizes from $\delta t=0.5$ to $\delta t = 0.01$. 
In addition to the convergence with bond dimension, the difference between the estimate
of the eigenvalue from the left and right eigenvectors serves as an additional check on accuracy.

We find that with very modest computational resources ($D=3$, $\chi=100$), the SCGF easily converges
to approximately three significant digits, significantly greater than MF results. 
It is also clear, that unlike in quantum systems, where the variational principle prevents the ground state energy
from going below the exact ground state energy, our computed SCGF can go above and below the exact value. 
Also notable is that calculations in the jammed regime converge to more accurate results at a low bond dimension 
than those in the flowing region.
Without an initial set of sufficiently large time steps, we found that calculations in the jammed phase tend to converge
to local minima.

\begin{figure}
\centering
\includegraphics{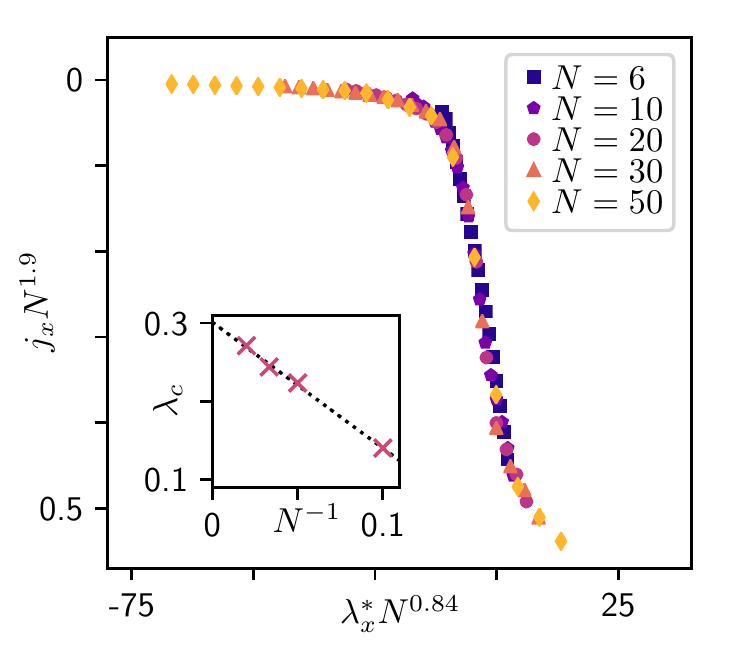}
\caption[Finite Size Scaling]{
Scaling plot of the transition between the flowing and jammed phases, showing the collapse of 
the per site horizontal current as a function of the reduced horizontal bias, $\lambda^*_x$. 
The inset plot shows a finite size extrapolation to estimate of the critical point $\lim_{N\to\infty}\lambda_c=0.30$, 
with $\lambda_c(N)$ by fitting a quadratic function to the three largest points in the susceptibility peaks 
for each $N$ in Figure \ref{fig:dpt}
}
\label{fig:fss}
\end{figure}

Last, we can perform a finite size scaling analysis of the observed transition to extract the critical exponents
in the thermodynamic limit.
Because the system sizes studied are limited to a linear dimension of  $N\le50$, the results retain
some finite-size error, though we expect that future work performing PEPS calculations on larger lattices,
possible because PEPS calculation costs grow linearly with system size,
or adapting infinite PEPS algorithms~\cite{phien2015infinite} could further refine these estimates.
The scaling relation for the per site horizontal current is
$j_x(\lambda^*_x,N)=N^d f(\lambda^*_xN^c)$,
where $d$ and $c$ are critical exponents, $f$ is the scaling function,
and $\lambda^*_x$ is analogous to a reduced temperature,
i.e. $\lambda^*_x=(\lambda-\lambda_c)/\lambda_c$. 
The inset of Figure \ref{fig:fss} shows a linear extrapolation of the location of the susceptibility peaks in 
Figure \ref{fig:dpt} to determine the critical point to be $\lim_{N\to\infty}\lambda_c=0.30$.
The critical parameters are then computed via numerical data collapse~\cite{bhattacharjee2001measure}, 
giving $d=-1.9\pm0.1$ and $c=0.84\pm0.1$ 
with Figure \ref{fig:fss} showing the resulting scaling plot, which displays good data collapse.


\section{\smallskip}
\noindent
{\textbf{\emph{Conclusions --}}}
We have provided the first insights into the dynamical phase behavior of the fully 2D ASEP,
finding evidence for a dynamical phase transition between a flowing and a jammed phase, 
as detected by a sharp change in the current in the horizontal and vertical directions.
We have also demonstrated how 2D tensor networks, in particular the PEPS ansatz, can be used
to compute large deviation functions in classical nonequilibrium systems,
characterize nonequilibrium phases, and obtain critical exponents.
This is a natural extension of the success of 1D tensor network methods in this field
and provides significant promise for the future use of TNs in coordination with LDT. 
Because numerical methods based on PEPS are relatively young, continued progress
is likely, and we expect such higher dimensional TNs to become standard tools
in the study of nonequilibrium classical statistical mechanics. 

\begin{acknowledgments}
\smallskip 
\noindent
{\textbf{\emph{Acknowledgments --}}}
This work was supported primarily by the US National Science Foundation via award no. 1665333.
PH was also supported by a NSF Graduate Research Fellowship under grant DGE-1745301 and an ARCS Foundation Award. 
\end{acknowledgments}
\bibliography{main}
\end{document}